\documentclass[aps,prd,longbibliography,twocolumn]{revtex4-2}
\usepackage[utf8]{inputenc}
\usepackage[dvipdfmx]{graphicx}
\usepackage{bm}
\usepackage {amsmath}
\usepackage{color}
\usepackage{ulem}
\usepackage{comment}
\begin{document}

\preprint{KUNS-2919, YITP-xx-xx, OCU-PHYS-557, AP-GR-178}

\title{Efficient search method of anomalous reflection by the central object 
in an EMRI system by future space gravitational wave detectors}
\author{Norichika Sago$^{a,b}$}
\author{Takahiro Tanaka$^{a,c}$}
\affiliation{$^a$Graduate School of Science, Kyoto University, Kyoto 606-8502, Japan}
\affiliation{$^b$Advanced Mathematical Institute, Osaka City University, Osaka 558-8585, Japan}
\affiliation{$^c$Center for Gravitational Physics, Yukawa Institute for Theoretical Physics, Kyoto University, Kyoto 606-8502, Japan}
\date{\today}

\begin{abstract}
In our previous work we investigated the effect of the hypothetical reflecting boundary near the black hole event horizon on the waveform from extreme mass-ratio inspirals (EMRIs). Even if the reflection efficiency is not extremely high, we found that a significant modification of the waveform can be expected. Then, the question is how to implement the search for this signature in the actual data analysis of future space gravitational wave antennas, such as LISA. In this paper we propose a simple but efficient method to detect the signature of the reflecting boundary. The interesting feature of the effect of the reflecting boundary on the orbital evolution of EMRIs is that the energy and angular momentum loss rates periodically oscillate in the frequency domain. The oscillation period is corresponding to the inverse time scale for the round trip of gravitational waves between the hypothetical boundary and the angular momentum barrier. We will show that this peculiar feature allows to detect the signature of the reflecting boundary without much additional computational cost. 
\end{abstract}

\maketitle

\section{Introduction}
Direct detection of gravitational waves (GWs)
\cite{LIGOScientific:2016aoc,LIGOScientific:2018mvr,LIGOScientific:2020ibl,LIGOScientific:2021djp} 
and direct imaging of a black hole (BH) shadow
\cite{EventHorizonTelescope:2019dse,EventHorizonTelescope:2022xnr}
have been realized recently. 
These achievements have opened a new window to test General Relativity by 
observing BHs.
As these observations are further developed, in near future they will bring us unprecedented
information on the nature of BHs and allow us to test the existence of the event horizon.
There are several works to seek different possibilities
that classical BHs are replaced with something else: 
for instance, horizonless compact objects like a boson star \cite{Liebling:2012fv}, 
gravastars \cite{Mazur:2001fv,Visser:2003ge}, 
firewall \cite{Almheiri:2012rt},
or quantum description of BHs motivated by string theory
\cite{Bekenstein:1974jk,Mukhanov:1986me,Skenderis:2008qn}.
In this paper, we refer to these alternative possibilities 
as Exotic Compact Objects (ECOs) to distinguish them from ordinary BHs predicted by General Relativity.

ECOs are expected to exhibit some different signatures from classical BHs 
(for example, Ref.~\cite{Cardoso:2019rvt} gives a comprehensive review about ECO)
and so far there are various proposals to distinguish BHs and ECOs by using the difference, 
for example, in the multipole structure
\cite{Wade:2013hoa,Krishnendu:2017shb,Kastha:2018bcr,Krishnendu:2019tjp},
in the tidal response (heating and deformability)
\cite{Cardoso:2017cfl,Sennett:2017etc,Agullo:2020hxe,Johnson-McDaniel:2018uvs,Narikawa:2021pak}.
In addition, the modification of the ringdown GWs in the post-merger phase of compact binary coalescences
(CBCs) and the subsequent echoes \cite{Cardoso:2016rao,Cardoso:2016oxy,Cardoso:2019apo} induced by
a reflecting inner boundary
are attracting attention, motivated by the claim that a possible signature of echoes
is found in the data after the binary black hole merger events observed by LIGO/Virgo
\cite{Abedi:2016hgu,Abedi:2017isz,Abedi:2020sgg}.
Although most of the follow-up analysis
\cite{Ashton:2016xff,Westerweck:2017hus,Nielsen:2018lkf,Lo:2018sep,Uchikata:2019frs,Tsang:2019zra,Abbott:2020jks,LIGOScientific:2021sio}
concluded that there is no significant evidence for echoes in the current data,
methods used in those analyses are different and they do not deny
the result in \cite{Abedi:2016hgu,Abedi:2017isz,Abedi:2020sgg,Abedi:2020ujo} directly. 
Ref.~\cite{Uchikata:2019frs} shows that some feature 
that cannot be explained by the simple detector noise is found by 
some analysis methods.
Furthermore, since GW echoes (if they are observed) can bring us rich information on ECOs,
the construction of the waveforms and analysis methods are still actively studied for
the future observation with higher sensitivity
\cite{Nakano:2017fvh,Mark:2017dnq,Maselli:2017tfq,Conklin:2019fcs,Maggio:2019zyv,Micchi:2019yze}.

Also there are several works on the signatures of ECOs composing extreme mass
ratio inspirals (EMRIs) in the tidal heating 
\cite{Maselli:2017cmm,Datta:2019euh,Datta:2020rvo,Datta:2019epe},
the tidal deformability \cite{Pani:2019cyc,Guo:2019sns,Datta:2021hvm}, 
and the area quantization \cite{Agullo:2020hxe,Datta:2021row}.
Unlike CBCs observed by ground-based detectors,
it is difficult to see the effect of the replacement of the inner boundary on
the post-merger GWs because of the small signal-to-noise ratio.
Fortunately, in EMRI case, we will observe many cycle of GW oscillations during the
observation period. Hence,
we can observe the modification of the orbital evolution and the corresponding GW phase
of EMRIs induced by the reflected waves by the inner boundary
\cite{Maselli:2017cmm,Guo:2019sns,Cardoso:2019nis,Sago:2021iku,Maggio:2021uge}.

In our previous paper \cite{Sago:2021iku}, we found that the modification to the GW phase due to the reflection 
is composed of two parts, the oscillatory part and the non-oscillatory one. 
The non-oscillatory part can be mostly absorbed by the shift of the ordinary binary parameters 
(the coalescence time, the overall phase, and two masses of the EMRI binary), 
while the oscillatory part is an unique definitive feature for the case with a reflecting near-horizon 
boundary.

In general, to test the possibility of models that go beyond the standard prediction based 
on General Relativity, a dedicated event search with the modified waveform templates is required.  
Usually, the calculation of the modified waveform with additional 
dimensions of the parameter space to search increases the computational cost significantly.
In this work, we propose a simple analysis method valid for this particular case 
to find the periodic oscillation of the GW amplitude in the Fourier space.  
Our method is based on the standard matched filtering technique with the ordinary (unmodified)
EMRI waveforms. Therefore, our method does not require much additional cost in searching for 
the signature of the reflecting boundary.

This paper is organized as follows.
In Sec.\ref{sec:waveform}, we briefly review the gravitational perturbation
in EMRI system in which the central BH is replaced by an ECO, 
based on the black hole perturbation theory, and the
adiabatic waveforms used in the analysis of this paper.
In Sec.\ref{sec:search}, we examine the distinguishability between a BH and an ECO in EMRI system
by evaluating the match between the modified waveform and the original one. In addition,
based on the property of the match, we propose a simple method to search for the signature
of reflecting inner boundary in EMRIs.
Finally, we summarize the paper in Sec.\ref{sec:summary}.
Throughout this paper we adopt the geometrized units with $G=c=1$.

\section{Adiabatic evolution of EMRIs and the waveform} \label{sec:waveform}

\subsection{Formulation of gravitational perturbations for reflecting inner boundary}
The orbital evolution and the gravitational waves of EMRIs are often analyzed by
using the black hole perturbation theory. In this approach an EMRI is described as
a particle orbiting a Kerr black hole, which is referred as a "BH-EMRI" in this
paper. On the other hand, to investigate the effect of the hypothetical reflecting boundary
near the horizon, we replace the central BH in BH-EMRI system by an ECO with the same mass
and slightly larger radius than the BH horizon. We call this system as an "ECO-EMRI".
We assume that the background geometry outside the surface of the central ECO 
is described by Kerr metric and then the equation of perturbations in ECO-EMRI systems
is the same one as in BH-EMRI, while the perturbations on the surface satisfy a reflecting
boundary condition, not the pure ingoing one.

In the previous works \cite{Sago:2021iku,Maggio:2021uge}, 
how to calculate the perturbations in ECO-EMRI system has been already
discussed. In this subsection, we briefly review the formulation given
in our paper \cite{Sago:2021iku}.

The linear gravitational perturbation in Kerr spacetime
is described by the Teukolsky variables, $\Psi_s$ ($\Psi_s$ is defined by the Weyl
scalars as $\Psi_2=\psi_0$ and $\Psi_{-2}=(r-ia\cos\theta)^{-4}\psi_4$, respectively)
\cite{Teukolsky:1973ha}.
We decompose $\Psi_s$ in the form of 
\begin{equation}
\Psi_s = \sum_{lm} \int d\omega 
R_{lm\omega}(r) S_{lm}^{a\omega}(\theta) e^{-i\omega t+im\varphi}\,,
\label{eq:Psi_s}
\end{equation}
where $S_{lm}^{a\omega}$ is the spheroidal harmonics and $R_{lm\omega}$ satisfies
the radial Teukolsky equation
\begin{equation}
\Delta^{-s} \frac{d}{dr}\left( \Delta^{s+1}\frac{dR_{lm\omega}}{dr} \right)
+ V_{lm\omega}(r)R_{lm\omega}(r) = T_{lm\omega}(r)\,
\label{eq:radial-Teukolsky}
\end{equation}
with $\Delta=r^2-2Mr+a^2$.
The expressions of $V_{lm\omega}(r)$ and $T_{lm\omega}(r)$ are given in Ref.~\cite{Sasaki:2003xr}, 
for example.
We consider only the $(l,|m|)=(2,2)$ modes in this paper. 
From here on, we suppress the subscripts of the functions, $(l, m, \omega)$,
for brevity.

To construct the Green's function,
we prepare two homogeneous solutions of
Eq.~(\ref{eq:radial-Teukolsky}), which satisfy the asymptotic boundary conditions, 
\begin{align}
&R^\textrm{in}\!=\!\left\{\!\!
\begin{array}{ll}
r^{-1}e^{-i \omega r^*}
+{\cal R} r^{-1-2s}e^{i \omega r^*}\!, & \mbox{for}~ r^*\to +\infty\,,\cr
{\cal T} \Delta^{-s}e^{- i k r^*}\!, & \mbox{for}~ r^*\to -\infty\,,
\end{array}
\right. \label{eq:in-going-R} \\
&R^\textrm{up}=\left\{
\begin{array}{ll}
\tilde{\cal T}r^{-1-2s}e^{i \omega r^*}\,,\quad & \mbox{for}~ r^*\to +\infty\,,\cr
e^{i k r^*}+\tilde {\cal R} \Delta^{-s}e^{- i k r^*}\,,\quad & \mbox{for}~ r^*\to -\infty\,.
\end{array}
\right. \label{eq:up-coming-R}
\end{align}
where $dr^*/dr=(r^2+a^2)/\Delta$, $k=\omega -m\Omega_H$, $\Omega_H=a/2Mr_+$, 
and $r_+=M+\sqrt{M^2-a^2}$. $R^\textrm{in}$ satisfies the purely ingoing boundary
condition at the horizon, while $R^\textrm{up}$ satisfies the purely outgoing
boundary condition at infinity.
The coefficients, $\mathcal{R}$ and $\mathcal{T}$, corresponds to the reflection 
and transmission for the incident ingoing wave from infinity.
Similarly, $\tilde{\mathcal{R}}$ and $\tilde{\mathcal{T}}$ are the reflection and
transmission coefficients for the outgoing incident wave from the horizon. 

In the BH-EMRI case, we can construct the Green's function that satisfies the purely ingoing
condition on the horizon and the purely outgoing condition at infinity by using
the above two homogeneous solutions, 
\begin{align}
G(r,r') =& \frac{1}{W(R^\textrm{in},R^\textrm{up})}
\left\{ R^\textrm{in}(r)R^\textrm{up}(r')\theta(r'-r) \right. \cr
&\left. + R^\textrm{up}(r)R^\textrm{in}(r')\theta(r-r') \right\},
\label{eq:Green-fnc}
\end{align}
where $W(R^\textrm{in},R^\textrm{up})$ is the Wronskian and $\theta(r)$ is the
Heaviside step function.

To obtain the Green's function for the ECO-EMRI case, we introduce another
homogeneous solution,
\begin{align}
\tilde{R}^\textrm{in} \equiv&
R^\textrm{in} + \beta R^\textrm{up} \cr
\simeq& \beta e^{ikr^*} 
+ (\mathcal{T} + \beta\tilde{\mathcal{R}}) \Delta^{-s} e^{-ikr^*}
\quad \mbox{for}~ r^*\to r_\textrm{b}^*,
\end{align}
where $r_\textrm{b}^*$ is the value of $r^*$ on the surface of the ECO.
The coefficient $\beta$ is related to the reflection rate on the reflecting boundary,
$R_\textrm{b}$, as
\begin{equation}
\beta = \mathcal{T} R_\textrm{b} \frac{\epsilon_-}{\epsilon_+}
   \left(1-\tilde{\mathcal{R}}R_\textrm{b}\frac{\epsilon_-}{\epsilon_+}\right)^{-1}\,.
   \label{eq:beta}
\end{equation}
The factors $\epsilon_\pm$ appear in the formulas of the energy fluxes in terms of the
asymptotic amplitudes of GWs (see below).
The explicit expressions are given by~\cite{Teukolsky:1974yv}
\begin{align}
\epsilon_{+}^2 =&
\frac{\omega^3}{k(2Mr_+)^3(k^2+4\epsilon^2)}\,, \cr
\epsilon_{-}^2 =&
\frac{256(2Mr_+)^5 (k^2+4\epsilon^2) (k^2+16\epsilon^2) k \omega^3}
{|C_\textrm{SC}|^2}\,, \cr
\epsilon =& \frac{\sqrt{M^2-a^2}}{4Mr_+}, \cr
|C_\textrm{SC}|^2 =&
\left[ (\lambda+2)^2 + 4a\omega m - 4a^2\omega^2 \right] \cr
&\times
\left[ \lambda^2 + 36a\omega m - 36a^2\omega^2 \right] 
+ 144 \omega^2 ( M^2  - a^2 )  \cr
&+ 48 a\omega (2\lambda + 3) ( 2a\omega - m )\,,
\end{align}
where $\lambda$ is the separation constant between the radial and angular 
Teukolsky equations.
The Green's function for the reflecting boundary condition can
be constructed by replacing $R^\textrm{in}$ in Eq.~(\ref{eq:Green-fnc}) with
$\tilde{R}^\textrm{in}$.

Once we construct the Green's function, we can obtain the solution of 
Eq.~(\ref{eq:radial-Teukolsky}) and extract the asymptotic forms at 
$r^*\to \pm\infty$.
For the BH-EMRI case, we would have
\begin{align}
R =& \int_{-\infty}^{\infty} G(r,r')T(r')dr' \cr
=& \left\{
\begin{array}{ll}
Z^\infty \tilde{\cal T} r^{-1-2s}e^{i \omega r^*}\,,\quad & \mbox{for}~ r^*\to +\infty\,,\cr
Z^H {\cal T} \Delta^{-s}e^{- i k r^*}\,,\quad & \mbox{for}~ r^*\to -\infty\,, 
\end{array}
\right.
\end{align}
with the coefficients $Z^\infty$ and $Z^H$ given by
\begin{equation}
Z^{\infty/H}  = \frac{1}{W(R^\textrm{in},R^\textrm{up})}\int T(r) R^\textrm{in/up}(r) dr\,.
    \label{eq:asymptotic_amplitude}
\end{equation}
By using the asymptotic forms, we can derive the energy fluxes of the GWs 
at infinity and on the horizon as
\begin{equation}
    F^{(\infty)}=\frac{|\tilde{\cal T}|^2|Z^{\infty}|^2}{4\pi\omega^2}\,,\qquad 
    F^{(H)}=\frac{\epsilon_-^2 |{\cal T}|^2 |Z^{H}|^2}{4\pi\omega^2}\,.
    \label{eq:flux}
\end{equation}

In a similar manner, we can express the asymptotic forms of the radial solution
for the ECO-EMRI case in terms of $Z^{\infty/H}$ as
\begin{align}
\tilde{R} =&
\left\{
\begin{array}{l}
\left( Z^\infty + \beta Z^H \right) \tilde{\cal T} r^{-1-2s}e^{i \omega r^*}\,,
\quad \mbox{for}~ r^*\to +\infty\,,\cr
Z^H \left( {\cal T} + \beta \tilde{\mathcal{R}} \right) \Delta^{-s}e^{- i k r^*}
+ \beta Z^H e^{i k r^*}\,,\cr 
\hspace{4cm} \mbox{for}~ r^*\to -\infty\,, 
\end{array}
\right.
\end{align}
and the corresponding fluxes as
\begin{align}
F^{(\infty)}_\textrm{mod} =&
\frac{|\tilde {\cal T}|^2|(Z^{\infty}+\beta Z^H)|^2}{4\pi\omega^2}\,, \cr
F^{(H)}_\textrm{mod} =&
F^{(H-)}_\textrm{mod} - F^{(H+)}_\textrm{mod}
\label{eq:modified_flux}
\end{align}
with
\begin{align}
F^{(H-)}_\textrm{mod} =&
     \frac{\epsilon_-^2 |{\cal T}+\beta\tilde {\cal R}|^2
     |Z_{H}|^2}{4\pi\omega^2}\,, \cr
F^{(H+)}_\textrm{mod} =&
     \frac{\epsilon_+^2|\beta|^2|Z_{H}|^2}{4\pi\omega^2}\,,
\label{eq:modified_inner_flux}    
\end{align}
where $F^{(H +)}_\textrm{mod}$ and $F^{(H -)}_\textrm{mod}$ 
correspond to the fluxes of the incident wave to 
and the reflected wave by the inner boundary. With the aid of Eq.~(\ref{eq:beta}),
we would find that the ratio between these fluxes gives the reflection rate on the
inner boundary
\begin{equation}
\frac{F^{(H+)}_\textrm{mod}}{F^{(H-)}_\textrm{mod}}
= \frac{\epsilon_+^2 |\beta|^2}{\epsilon_-^2 |\mathcal{T}+\beta\tilde{\mathcal{R}}|^2}
= |R_\textrm{b}|^2.
\end{equation}

\subsection{Adiabatic waveform for quasi-circular equatorial orbit}
The specific energy and angular momentum of the satellite particle in a circular, 
equatorial orbit in Kerr spacetime are given by 
\cite{Bardeen:1972fi}
\begin{align}
E =& \frac{r_0^{3/2}-2Mr_0^{1/2} \pm aM^{1/2}}{r_0^{3/4}
\sqrt{r_0^{3/2}-3Mr_0^{1/2} \pm 2aM^{1/2}}}, \cr
L =& \pm \frac{M^{1/2}(r_0^2 \mp 2a M^{1/2} r_0^{1/2} + a^2)}
{r_0^{3/4}\sqrt{r_0^{3/2}-3Mr_0^{1/2} \pm 2aM^{1/2}}},
\end{align}
and the orbital angular velocity by
\begin{align}
    \Omega_\varphi = \pm \frac{M}{r_0^{3/2} \pm aM^{1/2}},
\end{align}
where $r_0$ is the orbital radius, the upper and lower sign refer to
co-rotating and counter-rotating orbits, respectively.

The gravitational perturbation induced by a point mass moving along a circular, equatorial
orbit has a discrete spectrum determined by $\Omega_\varphi$. The corresponding Teukolsky
variables in Eq.~(\ref{eq:Psi_s}) are given by
\begin{equation}
\Psi_s = \sum_{lm}
R_{lm\omega_m}(r) S_{lm}^{a\omega_m}(\theta) e^{-i\omega_m t+im\varphi}\,,
\label{eq:Psi_s_discrete}
\end{equation}
where $\omega_m=m\Omega_\varphi$.
Focusing on the $(l,m)=(2,2)$ mode of the gravitational wave at infinity,
the waveform can be read from the relation,
$\Psi_{-2}=(\ddot{h}_+ -i\ddot{h}_\times)/2$, and the asymptotic form of
$R_{lm\omega}$ as
\begin{align}
 h_+ - ih_\times =&
 -\frac{2}{r} \frac{Z_{22}^\infty}{4\pi^2 f_2^2}
 \frac{S_{22}^{a \omega_2}(\theta)}{\sqrt{2\pi}}
 e^{-2\pi if_2(t-r_*)+2i\varphi} \cr
 \equiv& A(f_2) e^{-i\phi_0(f_2;t)}\,,
\end{align}
where $f_m=m\Omega_\varphi/(2\pi)$ and
\begin{align}
A(f) \equiv& \frac{2}{r}\frac{|Z_{22}^\infty|}{4\pi^2 f^2}
  \frac{S_{22}^{a\omega}(\theta)}{\sqrt{2\pi}}\,,\cr
\phi_0(f;t) \equiv&  2\pi f(t- r_*) - 2\varphi + \pi - \arg(Z_{22}^\infty(f))\, .
\end{align}

The evolution of a circular equatorial EMRI is driven by the energy loss due to the GW radiation.
The orbital angular velocity secularly changes at a rate given by
\begin{equation}
\frac{d\Omega_\varphi}{dt} =
\frac{d\Omega_\varphi}{dr_0}
\left(\frac{dE}{dr_0}\right)^{-1} \dot{E}.
\end{equation}
Here $\dot{E}$ is the rate of the energy loss of the orbiting particle,
which is given by the total energy flux radiated by GWs with the aid of the balance argument.
The GW frequency $f_m$ changes secularly according to the change of $\Omega_\phi$
and the waveform at the leading order of the adiabatic approximation is given by
\begin{align}
 h(t) =& A(f_2(t)) e^{-i\phi(t)},\cr
 \phi(t) \equiv&  2\pi \left(\int^t f_2(t') dt'-  f_2 r_*\right) \cr
 &- 2\varphi + \pi - \arg(Z_{22}^\infty(f_2(t)))\, .
\end{align}
We perform the Fourier transformation to the waveform
\begin{align}
 \tilde{h}(f) =&
 \int dt A(f_2(t)) e^{-i\phi(t)} e^{2\pi i ft}
 \nonumber \\ \simeq&
\frac{A(f)}{\sqrt{\dot{f_2}(t(f))}} e^{i(\Psi(f)-\pi/4)}\,,
\label{eq:adiabatic-waveform}
\end{align}
where we use the stationary phase approximation to reach the expression in the second line
of Eq.(\ref{eq:adiabatic-waveform}), and define 
\[
\Psi(f) \equiv 2\pi ft(f) - \phi(t(f))\,, \quad
t(f) \equiv \int \frac{df}{\dot{f}_2}\,.
\]
Under the stationary phase approximation, the plus and cross modes for $f>0$
in the frequency domain are given as
\begin{equation}
\tilde{h}_+(f) = \frac{1}{2}\tilde{h}(f)\,, \quad
\tilde{h}_\times(f) = i\tilde{h}_+(f) = \frac{i}{2}\tilde{h}(f)\,.
\end{equation}
For later use, we introduce the normalized waveform
\begin{equation}
\tilde{s}_+(f) = \mathcal{N} \tilde{h}_+(f)\,,
\quad
\tilde{s}_\times(f) = \mathcal{N} \tilde{h}_\times(f)\,,
\end{equation}
with the normalization factor
\begin{equation}
 \mathcal{N}^{-2} = (h_+|h_+) = (h_\times|h_\times)
= \int_{0}^\infty df \frac{A(f)^2}{\dot{f}_2 S_n(f)}\,.
\end{equation}
Here the inner product is defined by
\begin{equation}
(g|h) \equiv 2 \int_{-\infty}^{\infty} \frac{\tilde{g}(f) \tilde{h}^*(f)}{S_n(f)} df\,,
\end{equation}
and we adopt the noise spectrum density, $S_n(f)$, of LISA taken from Ref.~\cite{Robson:2018ifk}.

\section{Search for the signature of reflecting boundary}  \label{sec:search}

\subsection{Significance of the modification of waveforms}
\begin{figure*}
\includegraphics[bb=0 0 410 279, width=0.45\linewidth]{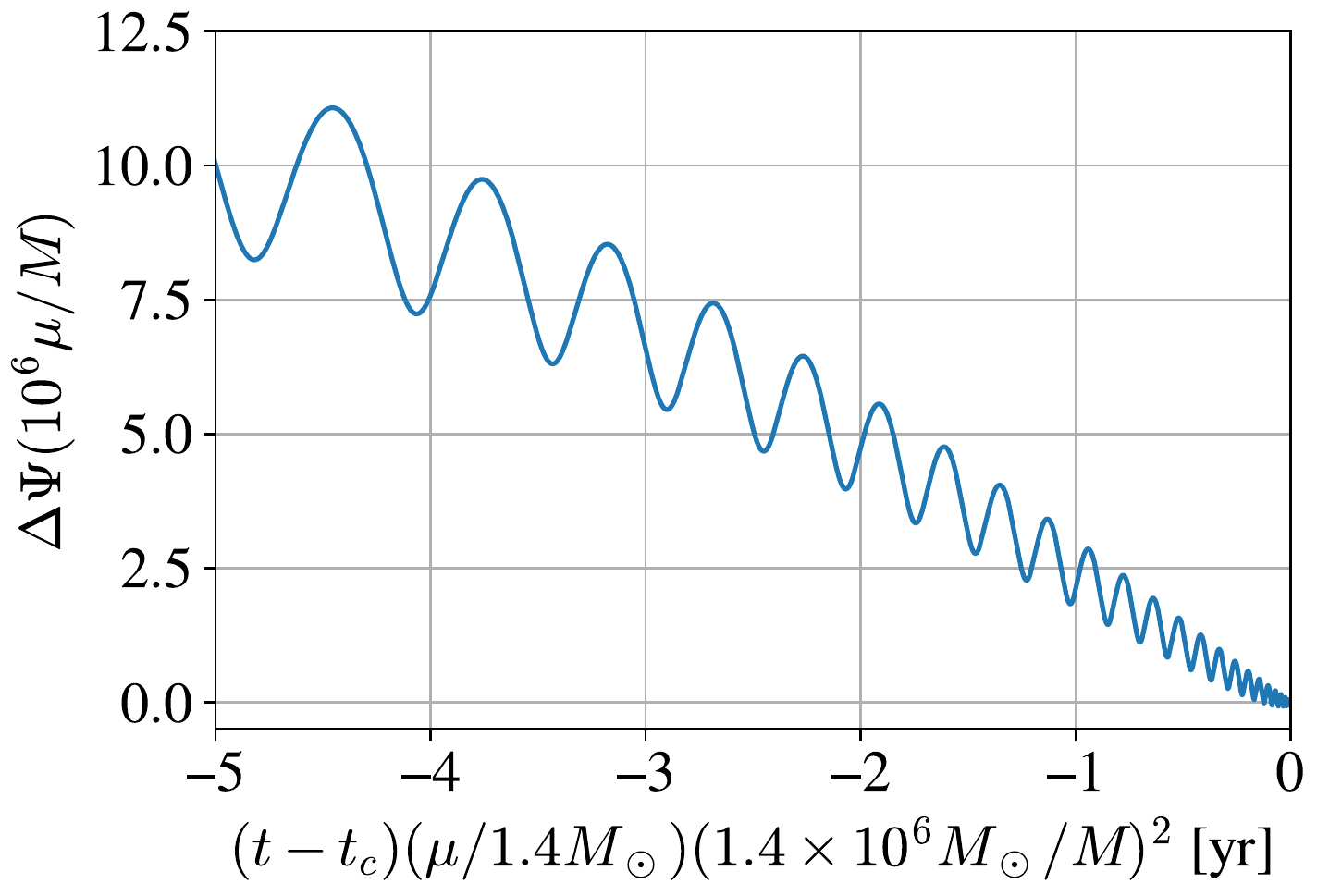}
\includegraphics[bb=0 0 398 279, width=0.45\linewidth]{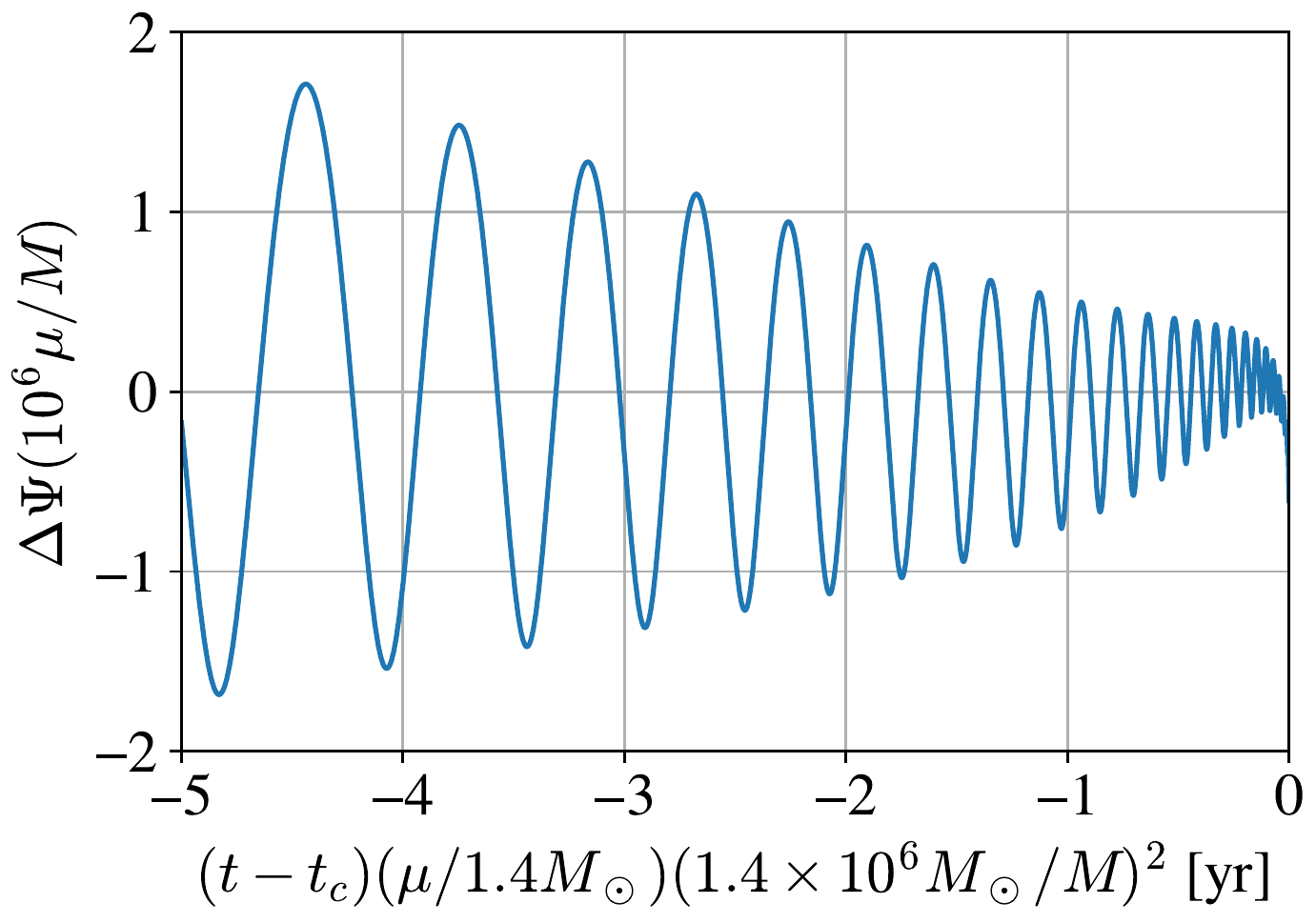}
 \caption{The modulation of the GW phase due to the reflecting boundary. In the left panel, we show
 the difference between the GW phase of a ECO-EMRI with 
 $(t_c,\mu,M)=(0,1.4M_\odot, 1.4\times10^6 M_\odot)$ 
 and that of a BH-EMRI with the same values of binary parameters. In the right panel, we show the
 difference between the two GW phases after we adjust the parameters of BH-EMRI waveform so that
 the match between two waveforms is maximized. In both plots, we fix the BH spin, the location of
 the reflecting boundary and the reflection rate as
 $(a,r_\textrm{b}^*,|R_\textrm{b}|^2)=(0.7M, -500M, 0.1)$.}
\label{fig:example-phase}
\end{figure*}

In our previous paper \cite{Sago:2021iku} we have discussed how a hypothetical reflecting boundary near
the horizon affects the evolution of EMRIs and the gravitational waveforms.
A key feature that we found is that the energy and angular momentum loss rates are modulated with 
the period corresponding to the inverse of the time scale for the round-trip of GWs between the 
reflecting boundary and the angular momentum barrier.
This feature is coincident with that found in \cite{Maggio:2021uge}.
In the following we would like to point out that there is an easy way to extract this modulation 
in the actual data analysis of EMRIs.  
The key idea is that this periodic modulation in the frequency domain is transferred to the multiple peaks 
in the time domain, after the inverse Fourier transformation, which is always done in the 
ordinary matched filtering to search for the best fit value of the coalescence time. 
These multiple peaks can be a clear evidence for the existence of the reflecting boundary. 
Another effect of the reflecting boundary on the phase evolution of EMRIs 
is to give a smooth component. However, as mentioned in the Introduction, 
this feature is difficult to detect
because the modification can be mostly absorbed by the shift of binary parameters 
(See the next paragraph and Fig.~\ref{fig:example-phase}).

To examine the detectability of the modulation in the GW phase,
we evaluate the match between the waveforms in the BH-EMRI case, $h$, and the
modified one in the ECO-EMRI case, $h_\textrm{mod}$. 
For simplicity, we consider a quasi-circular,
equatorial EMRI reaching the coalescence 
at the end and assume that the observation lasts for five years. 
We fix the parameters of the ECO-EMRI system to
$(t_c,\mu,M)=(0, 1.4M_\odot, 1.4\times 10^6 M_\odot)$, 
where $t_c$ is the time when the particle
reaches the ISCO radius. 
We search for the set of best-fit parameters, $(t_c, \mu, M)$,
that maximizes the match between two waveforms
\begin{equation}
{\cal M}(t_c, \mu, M) =
\frac{|(h_\textrm{mod}|h)|}
{\sqrt{(h_\textrm{mod}|h_\textrm{mod})(h|h)}}\,,
\label{eq:match}
\end{equation}
where we marginalize the phase of the EMRI waveform by taking the absolute value. 
Here we take the interval of the integration in the inner product as 
$f_\textrm{5yr} \le f \le f_\textrm{ISCO}$, where $f_\textrm{5yr}$ and $f_\textrm{ISCO}$
are the frequency five years before reaching the ISCO radius and that at the ISCO radius,
respectively.
We neglect the spin parameter in the search for the best fit parameters 
because its dependence is relatively small. Namely, the slowly varying phase shift 
depending on the frequency can be absorbed without using the degree of freedom of the 
black hole spin. 
In Fig.~\ref{fig:example-phase}, we show an example of the modulation of the GW phase
induced by the reflecting boundary, $\Delta\Psi$. The left panel shows the difference in GW phase
between the BH-EMRI and ECO-EMRI waveforms with the same binary parameters,
while the right panel shows the difference obtained by adjusting $(t_c,\mu,M)$
with a fixed value of $a$ so that the match is maximized. 
We can see that the non-oscillatory modulation is almost removed by the shift of
$(t_c,\mu,M)$ and that the spin parameter will not affect the maximization
in the case that the modification is moderate.

\begin{figure}
\includegraphics[bb=0 0 506 322, width=0.9\linewidth]{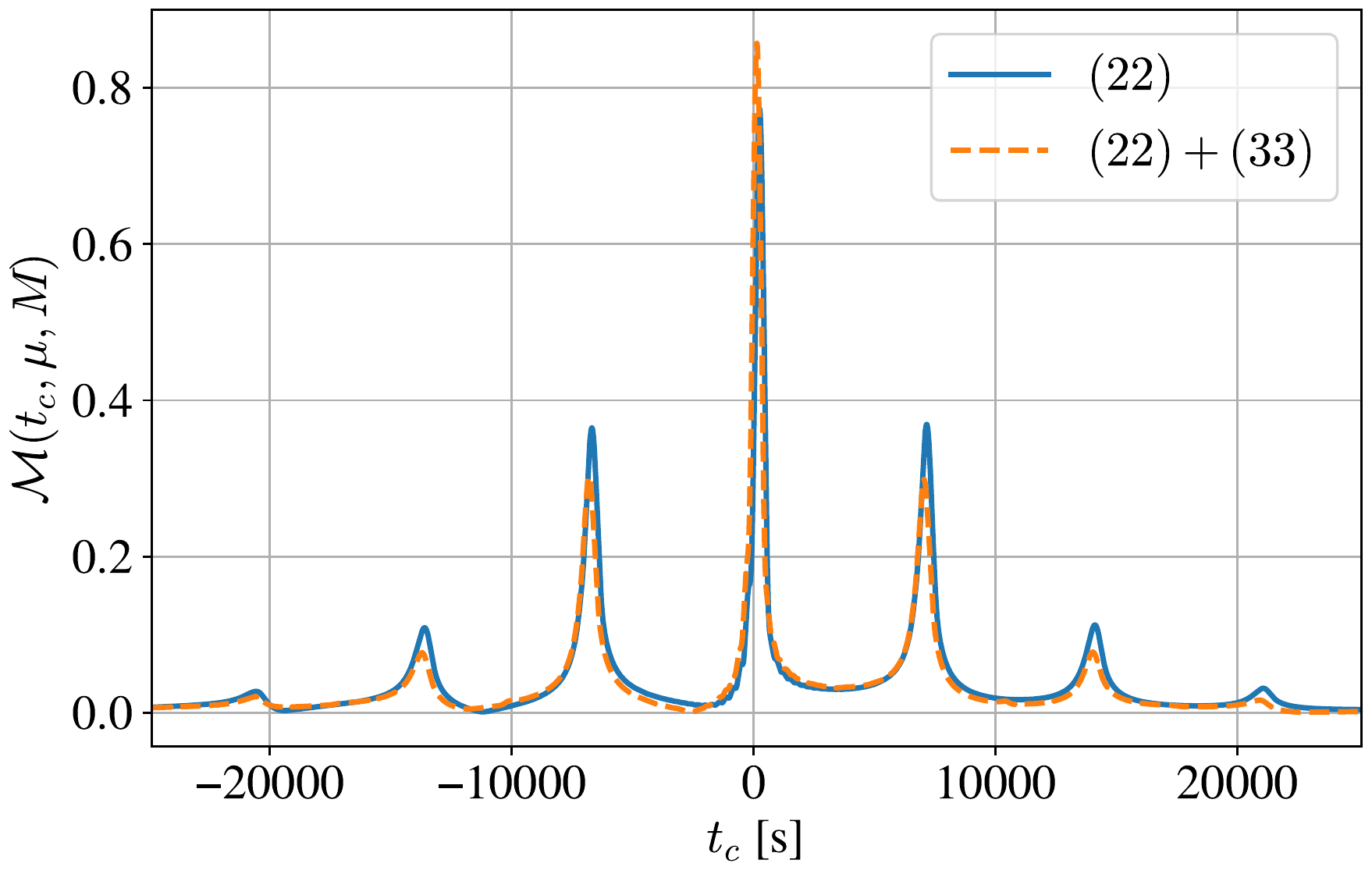}
 \caption{Match between an EMRI waveform and the modified one, when 
 the best-fit parameters are substituted into the mass parameters 
 of the search template waveform. We fix the BH spin, the location of
 the reflecting boundary and the reflection rate as
 $(a,r_\textrm{b}^*,|R_\textrm{b}|^2)=(0.7M, -500M, 0.1)$. The parameters
 to generate the modified waveform are set to $(t_c,\mu,M)=(0, 1.4M_\odot, 1.4\times 10^6 M_\cdot)$.
 The solid (blue) line corresponds to the case that only the flux of the $(l,|m|)=(2,2)$ modes
 is considered, while the dashed (orange) line shows the case that the $(l,|m|)=(3,3)$ modes
 are included in addition to the $(l,|m|)=(2,2)$ modes.}
\label{fig:Match_a07_sqR01}
\end{figure}

In Fig.~\ref{fig:Match_a07_sqR01}, we show the match with the best-fit mass
parameters as a function of $t_c$ for $a=0.7M$ and
$(r_\textrm{b}^*, |R_\textrm{b}|^2)=(-500M, 0.1)$. 
The largest peak corresponds to the best fit value of $t_c$.
We can find a series of damping peaks on both sides of the largest peak.
The value of the match at the highest peak is $0.773$, 
while the second peaks on the left and right sides
of the largest one are $0.365$ and $0.369$, respectively.
The interval between two successive peaks is corresponding to the round-trip time interval and 
is roughly estimated by
\begin{equation}
\Delta t_p = 2|r_\textrm{b}^*| \sim 7\times10^3\mbox{s}\,.
\end{equation}
These secondary peaks 
are caused by the modulation in the rate of change of the frequency due to the orbital evolution.
This example suggests the effectiveness of a simple test for the hypothesis of the reflective
boundary condition near the horizon just by searching for the secondary peaks in
the sequence of the correlations between the data and the template for
BH-EMRI system with respect to the coalescence time $t_c$.
In Sec.~\ref{sec:strategy}, we will propose an analysis method to search for the secondary peaks.

The $(l,|m|)=(2,2)$ modes are dominant in the energy flux in general and 
the correction due to the other modes is subdominant. Therefore we expect 
that the effect of the higher modes on the oscillatory modulation will be 
a fraction. To verify this expectation, we include the next leading modes, 
$(l,|m|)=(3,3)$, in the energy flux and calculate $\mathcal{M}(t_c,\mu,M)$
in the same manner. In Fig.~\ref{fig:Match_a07_sqR01}, we also plot the result
by the orange line. Comparing two graphs, 
we find that the secondary peaks are slightly suppressed when we include 
the higher modes. This is because the amplitude of the oscillatory
modulation of the GW phase is inversely proportional to the rate of change
of the frequency, $df/dt$, which is increased by the contribution of $(l,|m|)=(3,3)$ modes,
as we showed in our previous paper (See Eq.(28) in 
Ref.~\cite{Sago:2021iku}). We can also find that there are tiny peaks between 
the successive large peaks, which are induced by the inclusion of the $(3,3)$
contribution (although they may be too small to find in the present figure at
a glance). Since these changes are very small, however, they affect neither 
the basic structure of the leading and the secondary peaks nor the analysis 
proposed in Sec.~\ref{sec:strategy} so much.
From the above, we consider only the $(l,|m|)=(2,2)$ modes for simplicity.

\begin{figure*}
\includegraphics[bb=0 0 386 290, width=0.45\linewidth]{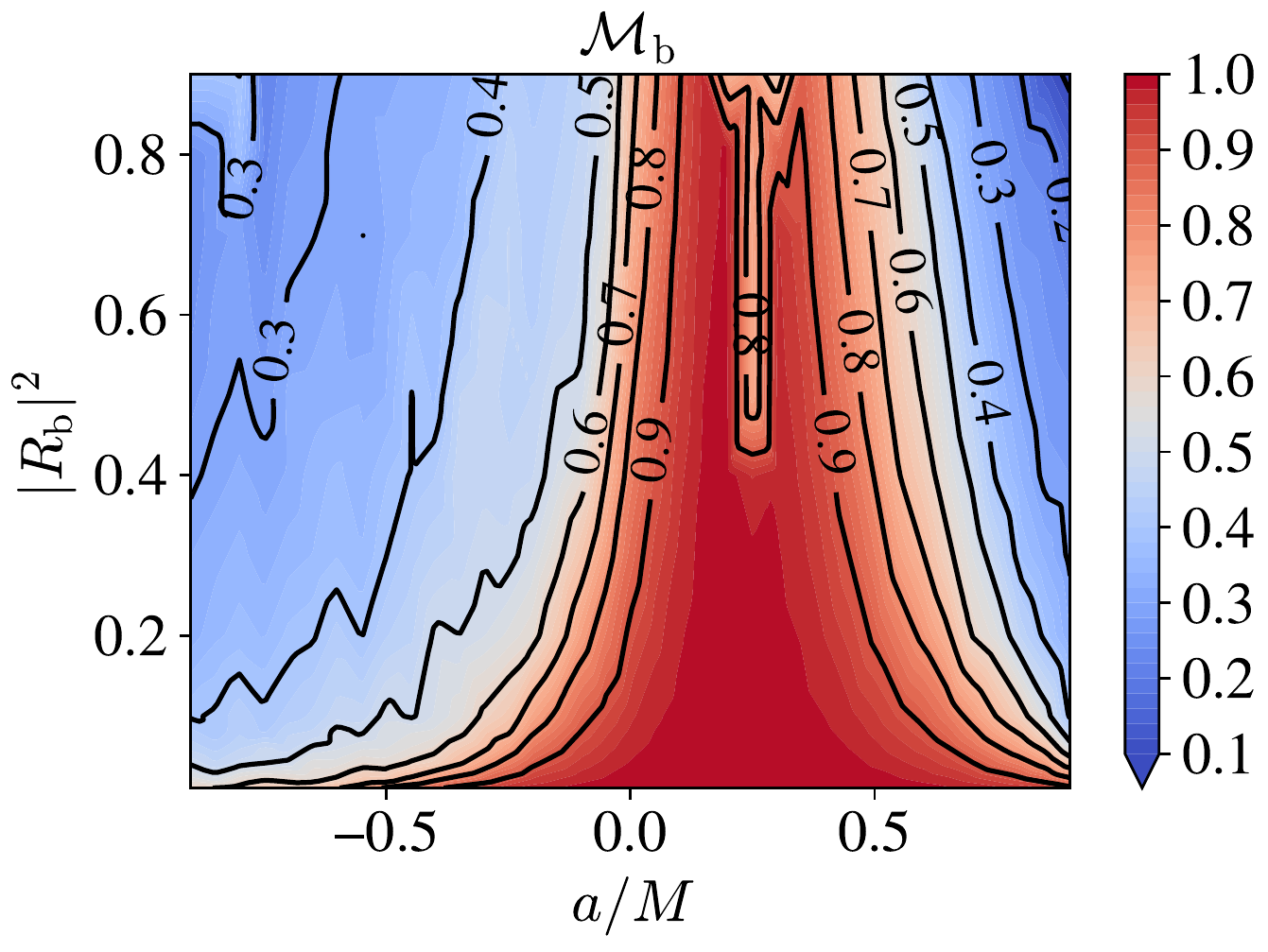}
\includegraphics[bb=0 0 387 291, width=0.45\linewidth]{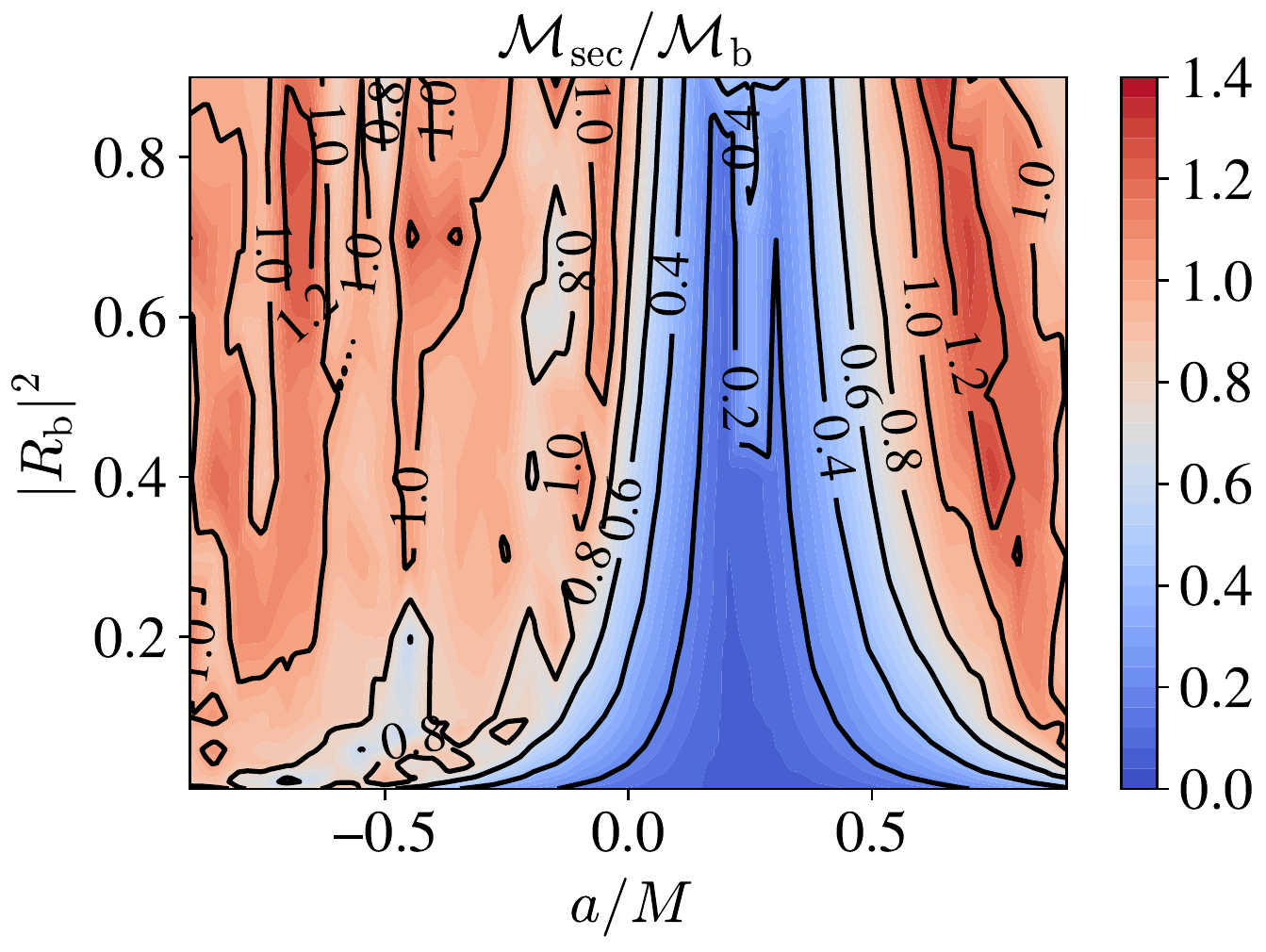}
 \caption{The dependence of the match on the spin and the reflection rate.
 Here, for convenience of drawing, we map co-rotating (counter-rotating) orbits 
 with respect to the BH's spin to positive (negative) values of $a/M$.
 In the left panel, we show the contour plot of $\mathcal{M}_\textrm{b}$ in the parameter space $(a, |R_\textrm{b}|^2)$.
 In the right panel, we show the contour plot of the ratio,
 $\mathcal{M}_\textrm{sec}/\mathcal{M}_\textrm{b}$.
 In both plots, we fix the location of the reflecting boundary and the binary
 parameters in the BH-EMRI waveform as $r_\textrm{b}^*=-500M$ and
 $(t_c,\mu,M)=(0, 1.4M_\odot, 1.4\times10^6 M_\odot)$, respectively.}
\label{fig:Match-contour}
\end{figure*}

In the left panel of Fig.~\ref{fig:Match-contour}, we show the contour plot 
of the match for the best-fit parameters, $\mathcal{M}_b$, 
in the parameter space $(a, |R_\textrm{b}|^2)$.
We find that the best-fit value of the match decreases when the black hole's spin and the
reflection rate of the boundary get large.
This can be understood as follows: The flux reflected by the boundary and the
corresponding modification of the orbital evolution increase in both cases. 
When the value of $a$ increases, the ingoing flux emitted by the orbiting particle 
gets larger, and then the reflected flux increases. As a result, the mismatch
between the waveforms in the BH-EMRI and the ECO-EMRI cases becomes large.

\begin{figure}
\includegraphics[bb=0 0 547 378, width=0.9\linewidth]{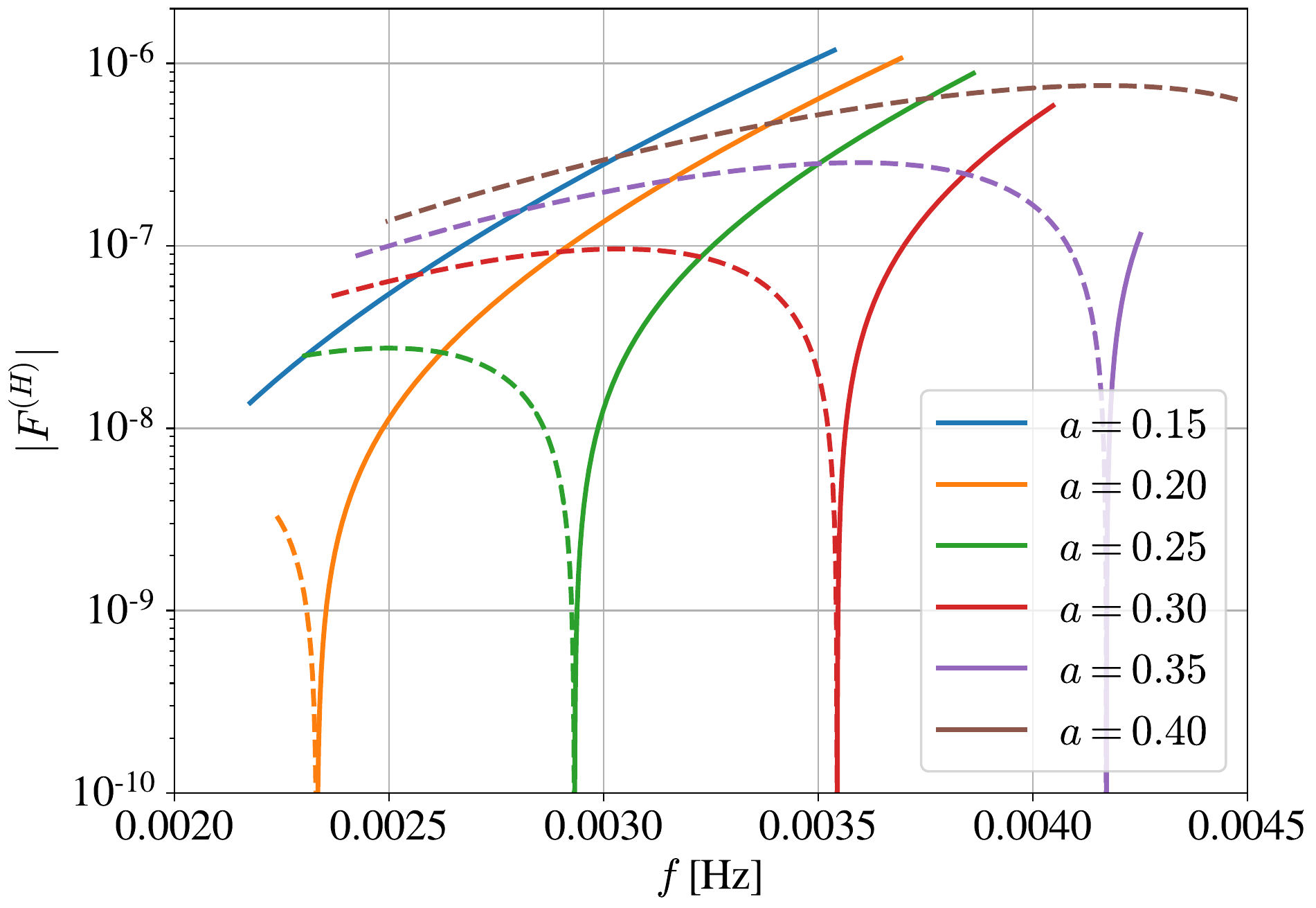}
 \caption{Dependence of the energy flux near the horizon, $F^{(H)}$, on
 the frequency for several values of the Kerr parameter.
 The masses of binary are fixed as $(\mu,M)=(1.4, 1.4\times 10^6)M_\odot$.
 Here, we plot the frequency range which is swept during the last five
 years before reaching the ISCO radius.
 The solid and dashed lines show the positive and negative
 (\textit{i.e.} superradiant) fluxes, respectively.}
\label{fig:FH_SRbound}
\end{figure}

We also find that the match in the counter-rotating case is relatively suppressed 
compared with that in the co-rotating case. The frequency evolution of counter-rotating EMRIs
is slower than co-rotating ones because the ISCO radius for the counter-rotating
case is larger than that of the co-rotating case. 
As a result, the oscillation in the energy flux also becomes slower. 
As the oscillation becomes slower, 
the amplitude of the phase modification becomes larger, if the relative 
amplitude of oscillation in the energy flux is the same
\footnote{The amplitude of the oscillatory modulation in GW phase is roughly
proportional to the oscillation period. See Eq.~(28) in \cite{Sago:2021iku}.}. 
This is the main reason 
why the match is lower in the counter-rotating case. 

In addition, the plot shows that $\mathcal{M}_\textrm{b}$ is suppressed around $a/M\sim 0.25$.
This is because the GW frequency crosses the threshold value for the 
superradiance during the five-year 
observation due to the orbital evolution of EMRI system. We show 
the plot of the flux near the horizon for various values of the Kerr parameter in the range
$0.15 \le a/M \le 0.4$ in Fig.~\ref{fig:FH_SRbound}.
We find that the sign of the flux changes during the observation
for $0.15 \lesssim a \lesssim 0.35$. The flip of the signature of the flux 
introduces an irregular modification to the GW phase, which is difficult to be absorbed 
by the shift of binary parameters.

In Fig.~\ref{fig:Match-contour}, we exclude the region of the parameter
space with (nearly) complete reflectivity or (nearly) extremal Kerr
parameter, in which there is some possibility of the ergoregion 
instability \cite{1973ZhETF..65....3S,1978CMaPh..63..243F,Maggio:2018ivz}. 
If the reflection rate is so high that rapidly rotating ECOs are 
unstable against the ergoregion instability within the observational
time scale of EMRIs, the angular momentum of such black holes should
be already extracted. 
The absence of a stochastic background of gravitational waves due to 
the spin down in the first observing run of LIGO already imposes constraints on ECOs
\cite{Barausse:2018vdb}.
Therefore, the probability of detecting gravitational waves from such 
rapidly rotating ECOs is expected to be quite low.

The decrease of the match leads to the loss of SNR, and then to the decrease of
the detection rate of the GW signal. For example, if $\mathcal{M}_\textrm{b}=0.4$,
the corresponding SNR falls to 40 \%, and then
the detection rate, roughly speaking, falls to $\sim 6.4 \%$ ($0.4^3\sim 0.064$).
This means that the matched filtering search with the BH-EMRI templates will not 
work efficiently if the modification due to the reflecting boundary is 
too large.
In this work, we are mainly interested in the parameter range of $(a,|R_\textrm{b}|^2)$ satisfying
$\mathcal{M}_\textrm{b} \gtrsim 0.4$. 

In addition to $\mathcal{M}_\textrm{b}$, we also calculate the root-sum-square value
of the matches at the two secondary peaks on both sides of the largest
peak, $\mathcal{M}_\textrm{sec}$.
In the right panel of Fig.~\ref{fig:Match-contour}, we show the contour plot
of the ratio, $\mathcal{M}_\textrm{sec}/\mathcal{M}_\textrm{b}$. The ratio gives us
the relative magnitude of the side-peak signal to the main signal.
There is a rough correlation that $\mathcal{M}_\textrm{sec}$ increases when
$\mathcal{M}_\textrm{b}$ decreases as expected
\footnote{
In this plot, we see some fine structure 
apparently in the region of $a<0$. Since the amplitude of the oscillatory
modulation in GW phase becomes large for the counter-rotating case, as mentioned above, the matched filtering used here
does not extract the information on the secondary peaks correctly.
Therefore the origin of the structure is not physical.}.

\begin{figure}
\includegraphics[bb=0 0 563 391, width=0.9\linewidth]{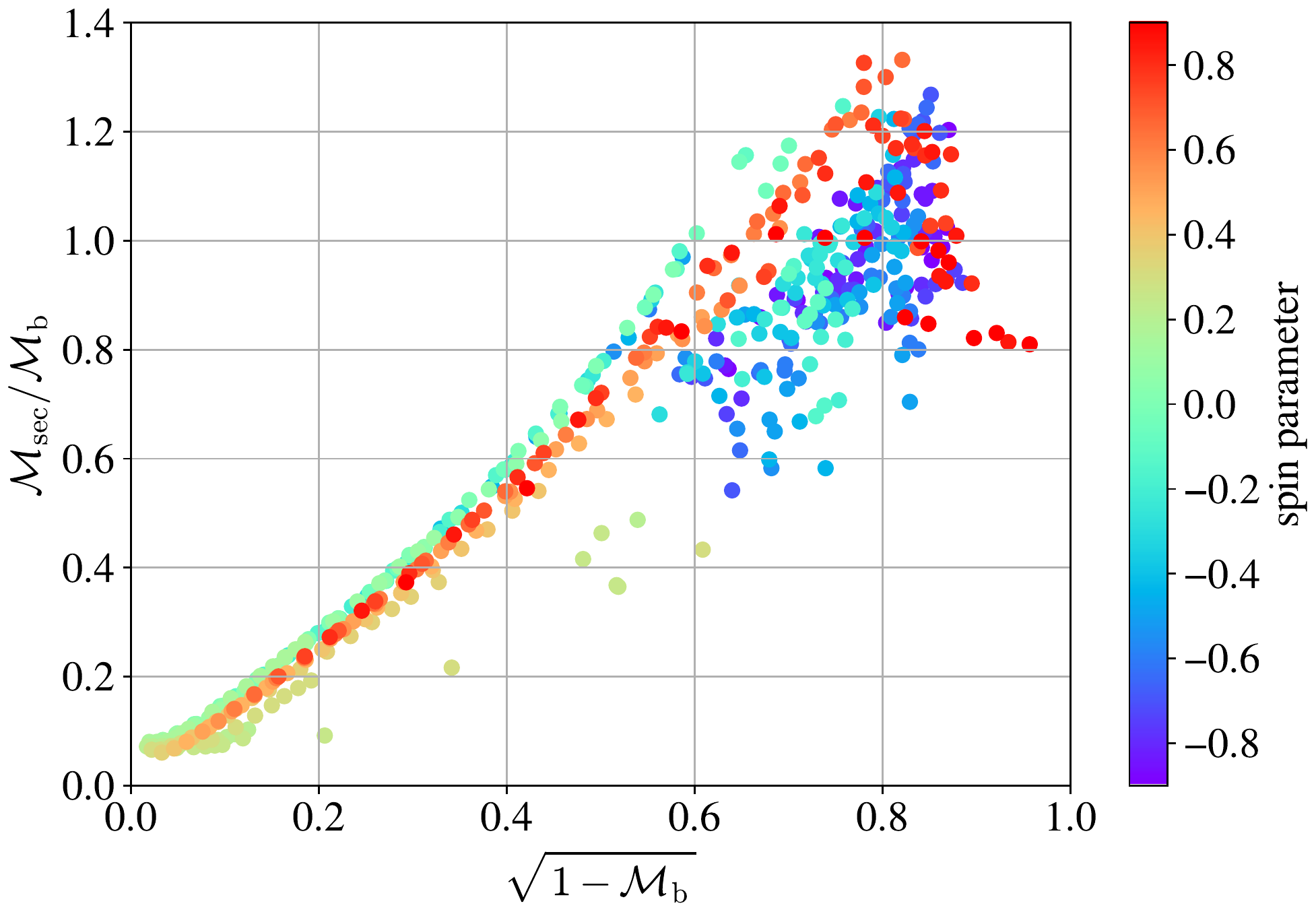}
 \caption{Relation between the best-fit peak and the side peaks in terms of the matches.
 The filled circles in the scatter plot show the data of 
 $(\mathcal{M}_\textrm{b}, \mathcal{M}_\textrm{sec})$
 calculated for several sets of $(a, |R_\textrm{b}|^2)$. The color is determined by the 
 spin parameter, $a$, shown in the color bar on the right hand side.
 Basically, the value of $|R_b|^2$ tends to be larger to the top right
of the graph. When $|R_b|^2$ get close to unity, however, 
the correlation between $\mathcal{M}_\textrm{b}$ and $\mathcal{M}_\textrm{sec}$
disappears.}
\label{fig:Mb_vs_Msec}
\end{figure}

In Fig.~\ref{fig:Mb_vs_Msec}, we show a scatter plot of
$(\mathcal{M}_\textrm{b}, \mathcal{M}_\textrm{sec}/\mathcal{M}_\textrm{b})$ 
for several sets of $(a,|R_\textrm{b}|^2)$.
The color of filled circles corresponds to the value of the spin parameter, $a/M$, as
shown in the color bar.
We find that most of the data points for $0.1 \lesssim \sqrt{1-\mathcal{M}_\textrm{b}} \lesssim 0.5$
($0.99 \gtrsim \mathcal{M}_\textrm{b} \gtrsim 0.75$) show a strong correlation between
$(\mathcal{M}_\textrm{b}, \mathcal{M}_\textrm{sec})$.
There are some points that largely deviate from the others, which 
belong to the exceptional region around
$a/M\sim 0.25$, and therefore do not show clean periodicity of the modulation
in the frequency domain. Apart from these exceptional cases, there is a good correlation. 
This correlation is lost for $\sqrt{1-\mathcal{M}_\textrm{b}}\gtrsim 0.5$, when the match 
is further reduced. Nevertheless, we find
that $\mathcal{M}_\textrm{sec}/\mathcal{M}_\textrm{b}\gtrsim 0.4$, as 
long as $\sqrt{1-\mathcal{M}_\textrm{b}}\gtrsim 0.32$, which corresponds to 
$\mathcal{M}_\textrm{b}\lesssim 0.9$. 
It should be noted that, in this plot (and also the right panel
of Fig.\ref{fig:Match-contour}),
the value of $\mathcal{M}_\textrm{sec}/\mathcal{M}_\textrm{b}$ is more than unity
in some cases. This is due to the definition of $\mathcal{M}_\textrm{sec}$, which is the root-sum-square
value of the matches at the two secondary peaks on both sides of the largest peak. 
For this reason, $\mathcal{M}_\textrm{sec}/\mathcal{M}_\textrm{b}$ can be larger 
than $1$ but should be less than $\sqrt{2}$.

\subsection{Search strategy for the secondary peaks in EMRI signals} \label{sec:strategy}

In the previous section and Fig.~\ref{fig:Match_a07_sqR01}, we showed that a series of
damping peaks on both sides of the largest peak emerge when we plot the match 
between the waveforms in the BH-EMRI and ECO-EMRI cases as a function of $t_c$.
In this subsection, based on this property, we propose an analysis method to search for 
the signature from a reflecting boundary by using the templates of BH-EMRIs.

\begin{figure}
\includegraphics[bb=0 0 585 254, width=0.9\linewidth]{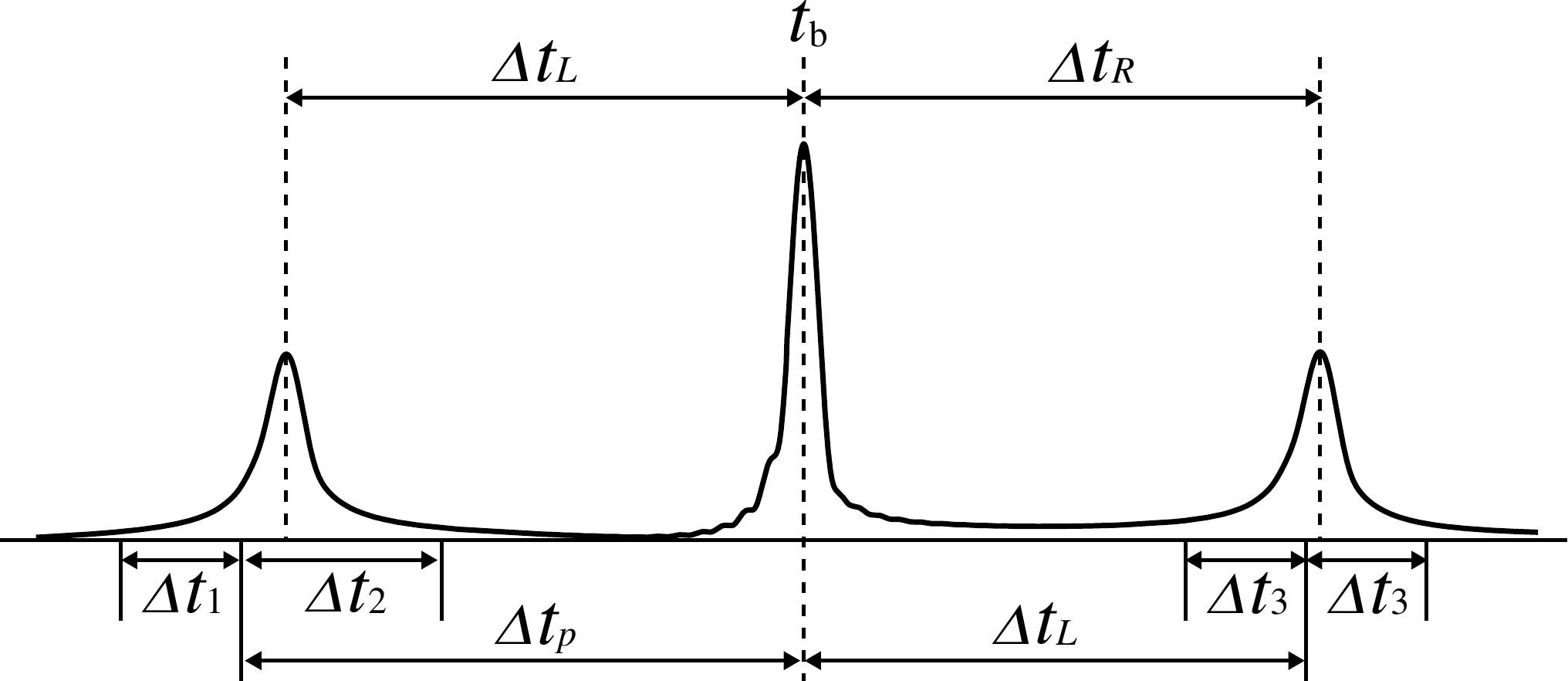}
 \caption{Image of search for the secondary peaks on both sides of the best-fitted
 peak in the domain of $t_c$.}
\label{fig:analysis}
\end{figure}

Suppose that a GW signal from a ECO-EMRI, $h_\textrm{mod}$, is included in the output 
of detector, $x=h_\textrm{mod}+n$, where $n$ is the noise.  
For simplicity, here we focus on the largest and the first secondary peaks on both
sides (See Fig.~\ref{fig:analysis}).
Our strategy is as follows:
\begin{enumerate}
\item 
Search for the best-fit parameters so that the signal-to-noise ratio (SNR) 
$\rho(t_c, \mu, M)=|(x|h(t_c,\mu,M)|$ is maximized,
by the matched filtering with the templates of BH-EMRI waveforms. 
Let $(t_\textrm{b}, \mu_\textrm{b}, M_\textrm{b})$ and $\rho_\textrm{b}$ 
be the best-fit values of the binary parameters
and the corresponding SNR, respectively.
\item
Create the time-series data of the SNR with the best-fit mass parameters,
$\rho(\Delta t_c)=|(x|h(t_c,\mu_\textrm{b},M_\textrm{b})|$, 
where $\Delta t_c \equiv t_c-t_\textrm{b}$.
\item
Search the left secondary peak of $\rho(\Delta t_c)$ in the interval of
$- \Delta t_p - \Delta t_1 \le \Delta t_c \le - \Delta t_p + \Delta t_2$,
where $\Delta t_1$ and $\Delta t_2$ are constants which determine the search region. 
In the current paper, we take 
$\Delta t_1=0.1\Delta t_p$ and $\Delta t_2=0.5\Delta t_p$.
Here, the position of the secondary peak on the left 
side of the largest peak is specified by $\Delta t_c=\Delta t_L$ .
\item
Search for the secondary peak of $\rho(\Delta t_c)$ on the right side of the largest peak 
in the interval of
$|\Delta t_L| - \Delta t_3 \le \Delta t_c \le |\Delta t_L| + \Delta t_3$, where
$\Delta t_3$ is a constant which determines the search window for the secondary peak on 
the right side. Here, we take $\Delta t_3=0.1\Delta t_p$. 
The position of the secondary peak on the right side of the largest peak is specified by 
$\Delta t_c = \Delta t_R$.
\item
Calculate the total SNR of the two secondary peaks defined by 
$\rho_\textrm{sec}^2=\rho^2(\Delta t_L)+\rho^2(\Delta t_R)$.
If $\rho_\textrm{sec}$ is larger than a certain predetermined threshold, record it
as a candidate of the detection of the modulation due to the reflecting boundary.
\end{enumerate}

We estimate the false alarm probability (FAP) for the search procedure mentioned above, 
assuming that the EMRI signal itself is already detected with a sufficiently large SNR using the 
BH-EMRI templates and the best-fit parameters in this case are known.  
For this purpose, we generate $10^6$ realizations of the stationary-Gaussian noise data,
which is 16,560 seconds long with the sampling rate of 0.1 Hz (the length corresponds 
to $2\Delta t_p + 2\Delta t_1 +2\Delta t_3 = 2.4\Delta t_p$), 
and apply the processes 2-5 to each data train.

\begin{figure}
\includegraphics[bb=0 0 404 274, width=0.9\linewidth]{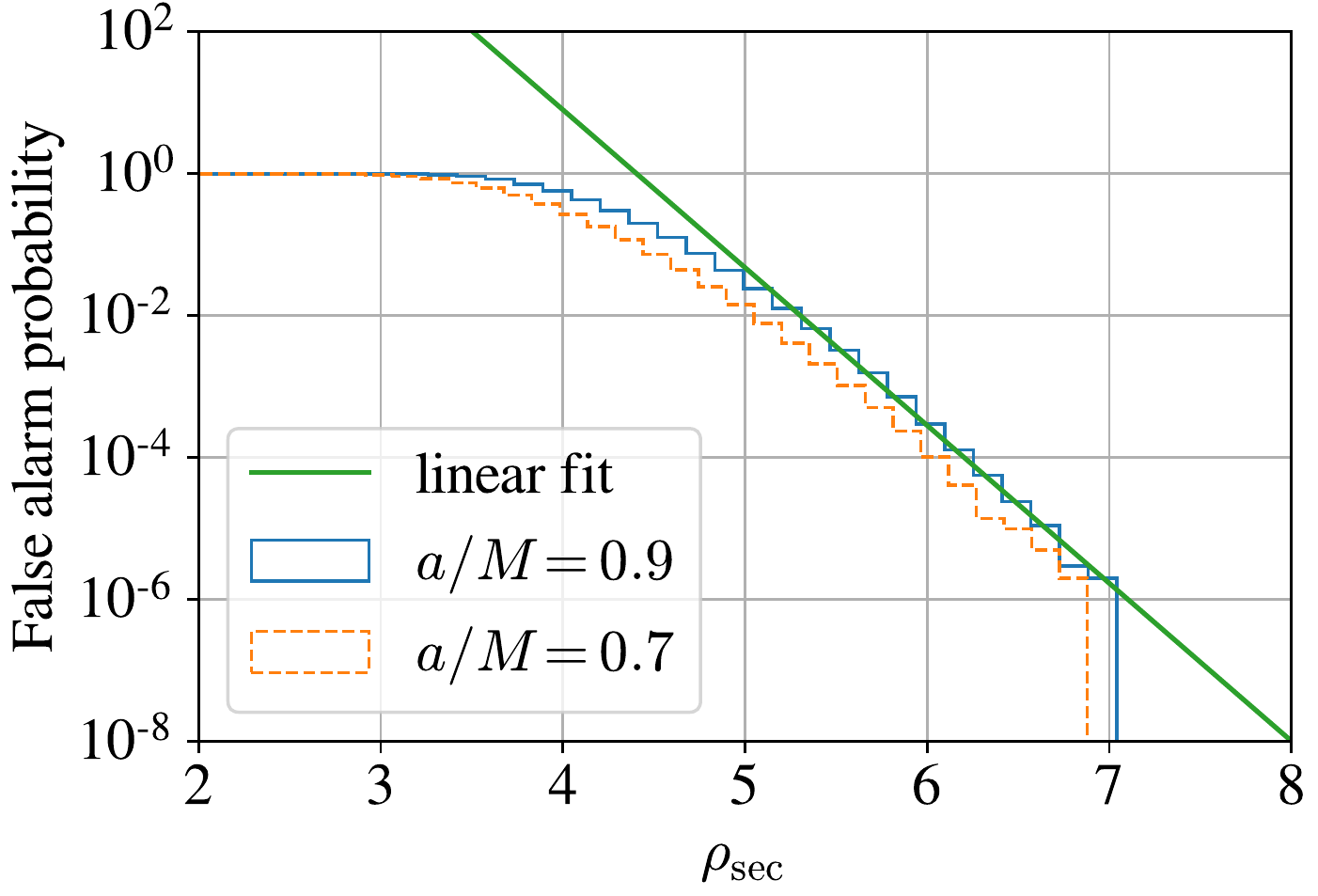}
 \caption{False alarm probability. The solid (blue) and dashed (orange) histograms 
 show the cumulative distributions of $\rho_\textrm{sec}$ normalized by the number 
 of realizations, $10^6$, for $a=0.9$ and $0.7$ respectively.
 The solid (green) line is the linear fit of the distribution for $\rho_\textrm{sec}>5$
 in the case of $a=0.9$.}
\label{fig:FAP}
\end{figure}

\begin{figure}
\includegraphics[bb=0 0 410 268, width=0.9\linewidth]{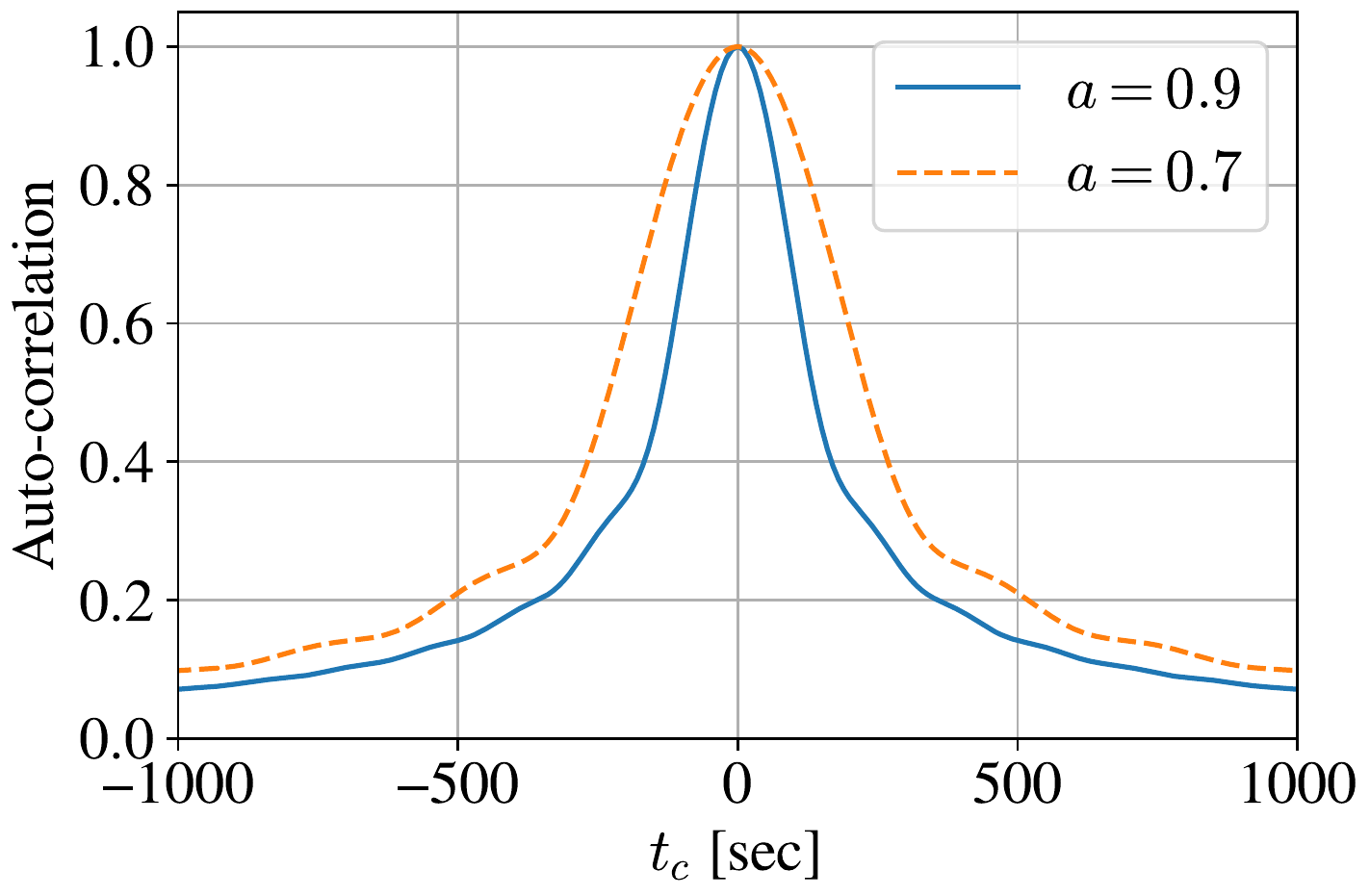}
 \caption{Auto-correlation of the waveform of BH-EMRI as a function of the difference 
 in the coalescence time. Here, we set the binary parameters
 of the waveform to $(\mu, M)=(1.4, 1.4\times 10^6)M_\odot$. 
 The solid (blue) and dashed (orange) curves show the auto-correlation 
 for $a=0.9$ and $0.7$, respectively.}
\label{fig:auto-correlation}
\end{figure}

In Fig.~\ref{fig:FAP}, we show the simulated FAP for $a=0.7$ and $0.9$.
We find the weak dependence of the FAP distribution on the value of $a$.
To understand the origin of this dependence, we calculate the auto-correlation of
the BH-EMRI waveforms with respect to the difference in the coalescence time 
shown in Fig.~\ref{fig:auto-correlation}. This plot shows
that the width of the peak of the auto-correlation gets narrower as $a$ increases.
This is because the ISCO frequency becomes higher and the cycle of GWs during 
the 5-year evolution increases for a larger value of $a$. 
The decrease of the width causes the increase of
the effective number of independent templates included in each data train, and
results in the increase of the FAP.
As is shown in the left panel of Fig.~\ref{fig:Match-contour}, the match with the BH-EMRI template, 
${\mathcal M}_\textrm{b}$, becomes significantly reduced in the high spin region. 
For the parameter region with ${\mathcal M}_\textrm{b}\lesssim 0.4$, the detectable 
event rate using the BH-EMRI template, roughly speaking, is reduced to less than 6.4\% compared
to the case without modification. 
Therefore, restricting our focus on the region with $a\lesssim 0.9$, 
we adopt the FAP distribution for $a=0.9M$ to determine a conservative threshold, 
which can be safely applied to all cases with $a\le 0.9$.
The linear fit of the FAP for $a=0.9$ is given by 
\begin{align}
\log(\textrm{FAP})& \simeq 21.0 - 5.02 \rho_\textrm{sec}\,. \label{eq:FAP-fit-a09}
\end{align}
For example, to achieve the FAP less than $10^{-8} (10^{-2})$, 
we should choose the threshold of SNR as 7.87 (5.11).

\begin{figure}
\includegraphics[bb=0 0 330 271, width=0.9\linewidth]{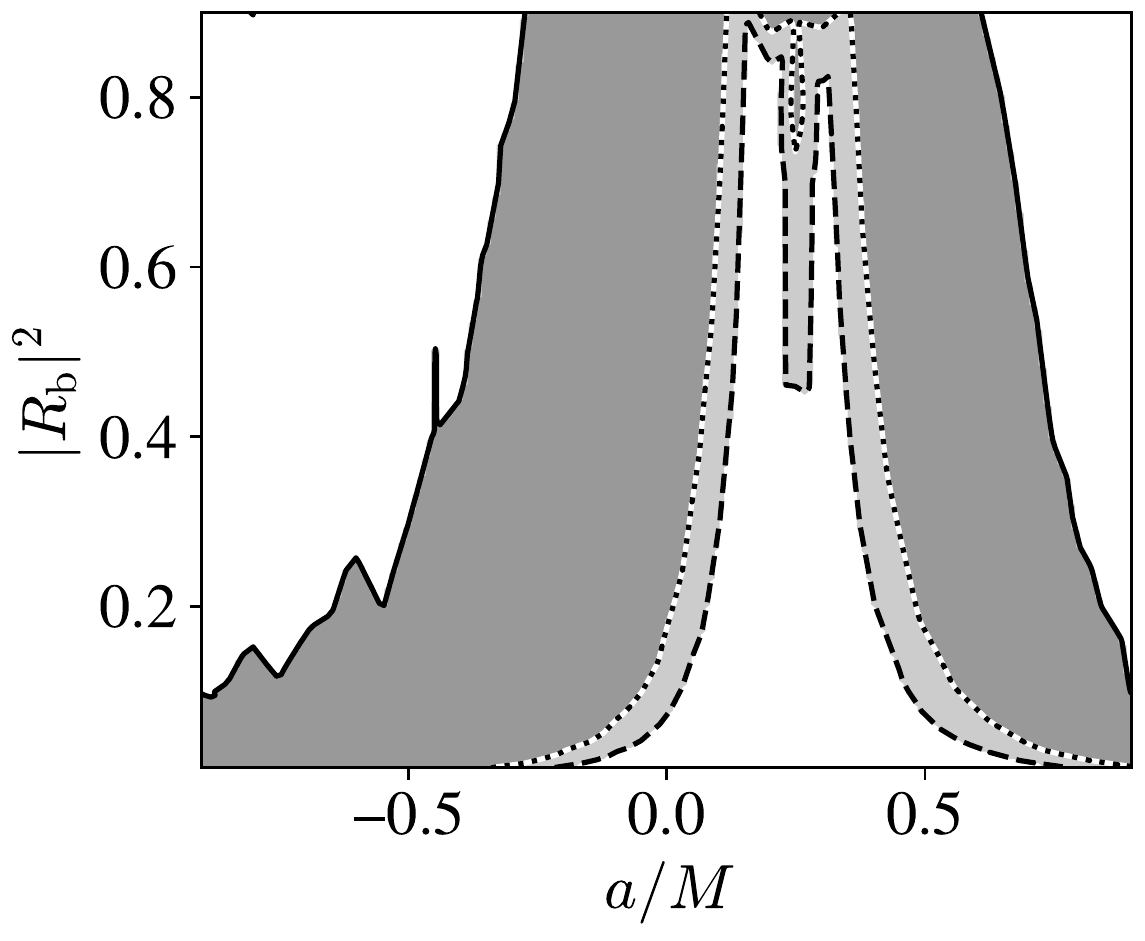}
 \caption{Applicable range in $(a,|R_\textrm{b}|^2)$ of the proposed analysis method.
 The solid, dashed, and dotted lines correspond to $\mathcal{M}_\textrm{b}=0.4$, 
 $\rho_\textrm{sec}=5.11$, and $\rho_\textrm{sec}=7.87$, respectively. 
 The shaded (both light and dark gray)
 region satisfies the condition that $\mathcal{M}_\textrm{b}>0.4$ 
 and $\rho_\textrm{sec}>5.11$.
 The dark gray region satisfies the condition that $\mathcal{M}_\textrm{b}>0.4$ and 
 $\rho_\textrm{sec}>7.87$.}
\label{fig:applicable-range}
\end{figure}

Assuming that the signal at the largest peak is observed with the SNR of $\rho_\textrm{b}$,
we can estimate the expected SNR of the side-peak signal, $\rho_\textrm{sec}$,
by multiplying $\rho_\textrm{b}$ by $\mathcal{M}_\textrm{sec}/\mathcal{M}_\textrm{b}$ given in
the right panel of Fig.~\ref{fig:Match-contour}.
As is shown in Fig.~\ref{fig:Mb_vs_Msec}, as long as $\mathcal{M}_\textrm{b}\lesssim 0.9$, 
we have $\mathcal{M}_\textrm{sec}/\mathcal{M}_\textrm{b}\gtrsim 0.4$, except for the case 
with $a\approx 0.25$.  
For the detection of EMRI system, we would require $\rho_\textrm{b}$ greater than 20 or so. 
Then, the SNR for the side band becomes 
larger than 8 for $\mathcal{M}_\textrm{b}\lesssim 0.9$. 
This means that the probability of the false detection of the secondary peaks is less than $10^{-8}$. 
Therefore, our method is applicable as long as the deviation from the BH-EMRI case is not 
too small ($\mathcal{M}_\textrm{b}\lesssim 0.9$) and the primary best fit signal can be detected 
using the BH-EMRI templates. 
We summarize the applicable range of the proposed detection method in the parameter space 
$(a,|R_\textrm{b}|^2)$ 
in Fig.~\ref{fig:applicable-range}.
In the light (dark) shaded region the false detection probability of the secondary peaks 
is $<10^{-8} (10^{-2})$ without significant reduction of the detection probability of the 
primary signal ($\mathcal{M}_\textrm{b}>0.4$).  

Our result in Fig.~\ref{fig:applicable-range} tells that the 
modification is detectable when the amplitude of the side band becomes large 
enough, which can be the case even for non-rotating case. One might consider
that it is inconsistent with the claim in \cite{Cardoso:2019nis}.
This is because the authors of this reference did not notice that 
the significant modification to the energy loss rate of EMRI occurs
even off resonance frequency. So, the effect was underestimated.

\section{Summary and discussion} \label{sec:summary}
In this work, we sought a new efficient method to detect the modification to the 
EMRI waveform due to the hypothetical 
reflecting boundary near the horizon.
We found that the match between the waveform for a BH-EMRI and that for an ECO-EMRI
has secondary peaks on both sides of the largest peak with respect to $t_c$.
The appearance of the secondary peaks originates from the oscillatory modulation in the phase of the ECO-EMRI waveform,
and we found that the square root of the sum of squared SNRs of the both side peaks 
is $\gtrsim 0.4$ relative to the SNR of the primary signal 
when the match of the ECO-EMRI waveform 
with the BH-EMRI templates, ${\cal M}_{\rm b}$, is in the range 
of ${\cal M}_{\rm b}\lesssim 0.9$. 
Namely, we have a significantly large signal of the modification as the secondary peaks, 
as long as the reduction of the SNR for the primary signal due to the modification exceeds 10\%. 
Therefore, if we can detect EMRIs with SNR greater than 20 using the BH-EMRI templates and 
${\cal M}_{\rm b}$ is less than 0.9 or so, 
we can expect to be able to detect the secondary peaks with the SNR greater than 8.
The probability that such a high SNR occurs is 
estimated to be less than $10^{-8}$, assuming the Gaussian noise. 

When the reflection rate is not large enough, ${\cal M}_{\rm b}$ becomes close to unity. 
In such cases the amplitudes of the secondary peaks are suppresses. 
By contrast, when the reflection rate is large or the black hole spin is large, 
the modification of the waveform can be too large, and ${\cal M}_{\rm b}$ is significantly reduced. 
When ${\cal M}_{\rm b}\lesssim 0.4$, the detection volume is reduced to less than 6.4\%, roughly speaking. 
Hence, the detection of such modified EMRI events by using the BH-EMRI templates becomes rare. 
However, the value of ${\cal M}_{\rm b}$ depends on the binary parameters, especially 
on the Kerr parameter of the central BH. 
For a moderate spin of the central BH, 
a larger value of the match can be realized even though the reflection rate is large. 

In this analysis, we used only the ordinary waveforms for BH-EMRI systems
as templates.
We need to implement a data analysis pipeline using BH-EMRI templates 
anyway to perform the EMRI event search. 
Therefore, the additional tasks and computational costs required for this search are minimal. 

In this work, we focused on some restricted situation to simplify the analysis.
First, we assumed that the reflection rate on the inner boundary is constant.
If the rate and the phase of the reflection coefficient sensitively depend on 
the GW frequency, the secondary peaks of the
SNR will become lower. Hence, the present analysis does not work. 
In such cases, a more concrete model of the reflective boundary such that 
allows us to predict the frequency dependence of the reflection coefficient 
would be necessary. 

Second, we restricted the EMRIs to the ones in quasi-circular orbits 
on the equatorial plane ($\theta=\pi/2$). 
It is expected that a certain fraction of EMRIs have a
large eccentricity and an inclination. 
Since the GW spectrum of such EMRIs is not monochromatic, the periodic 
modulation of the orbital phase will not appear as clear as the quasi-circular equatorial case.

Finally, we should mention the shortcoming by focusing only on 
the $(l,|m|)=(2,2)$ contribution to the energy loss
of the EMRI system to calculate its orbital evolution. 
Taking into account the higher harmonics
will introduce different periodicities in the modulation of the GW phase.
As we showed in Fig.~\ref{fig:Match_a07_sqR01},
the magnitude of their contribution is smaller than $(l,|m|)=(2,2)$ contribution
and it does not affect the basic strategy of our analysis, {\it i.e.}, searching for the secondary
peaks in matched filtering. To improve the accuracy of the parameter estimation, 
however, we should include the contribution of the higher harmonics to our analysis.
We leave this issue for future work.

\acknowledgments
This work is supported by JSPS Grant-in-Aid for Scientific Research JP17H06358 (and also JP17H06357), as a part of the innovative research area, ``Gravitational wave physics and astronomy: Genesis'', and also by JP20K03928. 
NS also acknowledges support from JSPS KAKENHI Grant No. JP21H01082.
\bibliography{bib_ECO_EMRI2}

\end{document}